\documentclass[twocolumn,pra,aps,superscriptaddress,10pt,longbibliography]{revtex4-2}
\usepackage{amsmath}
\usepackage{graphicx}
\usepackage[ bookmarks=true, colorlinks, linkcolor=blue, urlcolor=blue, citecolor=blue, plainpages=false, pdfpagelabels, final, breaklinks=true ]{hyperref}
\usepackage{color}

\begin{document}

\title{Discrete time crystal in an open optomechanical system}

\author{Dongni Chen}
 \affiliation{Key Laboratory of Frontiers in Theoretical Physics,
              Institute of Theoretical Physics,
              Chinese Academy of Sciences, Beijing 100190, China}
 \affiliation{School of Physical Sciences,
              University of Chinese Academy of Sciences,
              Beijing 100049, China}
							
\author{Zhengyang Peng}
 \affiliation{Key Laboratory of Frontiers in Theoretical Physics,
              Institute of Theoretical Physics,
              Chinese Academy of Sciences, Beijing 100190, China}
 \affiliation{School of Physical Sciences,
              University of Chinese Academy of Sciences,
              Beijing 100049, China}
							
\author{Jiahui Li}
\affiliation{Beijing Computational Science Research Center, Beijing 100193, People's Republic of China}

\author{Stefano Chesi}\email{stefano.chesi@csrc.ac.cn}
\affiliation{Beijing Computational Science Research Center, Beijing 100193, People's Republic of China}
\affiliation{Department of Physics, Beijing Normal University, Beijing 100875, People's Republic of China}

\author{Yingdan Wang}\email{yingdan.wang@itp.ac.cn}
 \affiliation{Key Laboratory of Frontiers in Theoretical Physics,
              Institute of Theoretical Physics,
              Chinese Academy of Sciences, Beijing 100190, China}
 \affiliation{School of Physical Sciences,
              University of Chinese Academy of Sciences,
              Beijing 100049, China}
\begin{abstract}
The spontaneous breaking of time translation symmetry in periodically driven Floquet systems can lead to a discrete time crystal. Here we study the occurrence of such dynamical phase in a driven-dissipative optomechanical system with two membranes in the middle. We find that, under certian conditions, the system can be mapped to an open Dicke model and realizes a superradiant-type phase transition. Furthermore, applying a suitable periodically modulated drive, the system dynamics exhibits a robust subharmonic oscillation persistent in the thermodynamic limit. 
\end{abstract}

\maketitle

\section{Introduction}
As an analogue to spatial crystals, Wilczek first proposed the idea of \textit{time crystals} in 2012~\cite{Wilczek2012_PRL109-160401}. Soon, it was pointed out that a system where continuous time invariance is spontaneously broken would naturally radiate energy into the environment, which conflicts with the principle of energy conservation~\cite{Bruno2013a}. Indeed, formal no-go theorems have shown that time crystals cannot exist in equilibrium~\cite{Bruno2013,Watanabe2015}. On the other hand, broken time translational invariance is still allowed under non-equilibrium conditions, where the concept of \textit{discrete time-crystals} (DTCs) has been proposed \cite{Khemani2016,Else2017,Yao2017}. A DTC is realized in a periodically driven system, with Hamiltonian satisfying $H(t)=H(t+T)$, and breaks discrete time translational symmetry, i.e., the period of the dynamics is a multiple of the driving period $T$~\cite{Else2020}. In a genuine DTC phase such spontaneously generated sub-harmonic response should be robust against parameter variations and persist to arbitrarily long times in the thermodynamic limit~\cite{Else2020,Yao2017,Fazio2017_PRB95-214307}. Experimentally, DTCs has been explored with trapped ions~\cite{Zhang2017_Nature543-7644}, vacancy-based quantum simulators~\cite{Choi2017_Nature543-221}, superfluid helium-3~\cite{Autti2018_PRL120-215301}, and spin NMR systems~\cite{Pal2018_PRL120-180602, Rovny2018_PRL120-180603}. Besides, various generalizations have been proposed theoretically, such as the realizations of a DTC in the Dicke model~\cite{Ueda2018_PRL120-040404, Bihui2019_NJP21-073028}, finite chains of Rydberg atoms~\cite{LaRocca2020_PRA101-013417}, in the presence of quasiperiodic spatial modulations~\cite{Pengfei2020}, or topological DTCs~\cite{Giergiel2019_NJP21-052003}. 

While most DTC realizations and proposals are based on interacting spin model, in the past decades optomechanical systems (where light interacts with motional degrees of freedom) have become one of the most promising platforms for exploring macroscopic quantum-mechanical behaviors and quantum information processing. This is due to their high coherence, the presence of an intrinsic nonlinear coupling, and the ability to couple in a versatile way to other quantum systems~\cite{Aspelmeyer2014,Barzanjeh2022}. A large variety of quantum engineering protocols have been proposed in optomechanics~\cite{Barzanjeh2022}. Among them, of special relevance here is a ``membrane in the middle'' setup realizing a Dicke-type phase transition~\cite{Mumford2015_AnnalenderPhysik527}. In that system, the mechanical mode and two cavity modes are mapped to the bosonic mode and collective spin of the Dicke model, respectively. However, cavity dissipation in optomechanical systems is normally much larger than the dissipation of the membranes. Hence, the mapping leads to a Dicke model where dissipation acts predominantly on the collective spin, unlike typical quantum-optics realization (where cavity dissipation dominates~\cite{Dimer2007_PRA75-013804,Nagy2010,Kirton2019_AdvancedQuantumTechnologies2-1970013}). Furthermore, applying the Schwinger’s spin-boson mapping leads to a collective decay of total angular momentum which differs from the more usual collective decay (see, e.g., Ref.~\cite{Gelhausen2017}) or individual spin decoherence~\cite{Kirton2017_PRL118-123602,Kirton2019_AdvancedQuantumTechnologies2-1970013,Reiter2020}.

Inspired by the above proposal, we consider here an alternative ``two membranes in the middle'' setup, which can realize a more typical Dicke-type phase transition. A main difference is that, in our model, light-matter degrees of freedom are not swapped by the mapping: the cavity and mechanical modes of the optomechanical setup correspond to the cavity mode and the collective spin of the Dicke model, respectively. Therefore, due to the much smaller decay rate of the membranes compared to cavity decay, the conservation of the atomic angular momentum (in the Dicke model) is a much better approximation. We show that the Dicke phase transition can be simulated in this optomechanical system with realistic parameters. Furthermore, we analyze the realization of a DTC phase, which for the Dicke model has been recently discussed in Refs.~\cite{Ueda2018_PRL120-040404, Bihui2019_NJP21-073028}. Unfortunately, the simple approach of pulsing on/off the effective coupling is not directly applicable to our system, due to the specific features of the optomechancal system and the mapping. Thus, we develop an alternative sequence of control pulses which can achieve an equivalent result.

The outline of our paper is as follows: In Sec.~\ref{Sec: the system} we introduced our ``two membranes in a cavity'' model and its mapping to the Dicke model. In Sec.~\ref{Sec: Quantum phase transition} we study the phase transition and phase diagram of this model. The validity of various approximations invoked in the mapping are also checked. In Sec.~\ref{Sec: time-crystal} the pulse sequence to realize the DTC is presented. We also provide discussions on various issues such as the choice of flipping time, the robustness of the DTC phase, the fate of the DTC in the deep quantum regime, and the influence of mechanical damping. Finally, we summarize our work in Sec.~\ref{Sec: Summary}. Some technical details are given in Appendices~\ref{Appendixa} and \ref{Appendixb}.

\section{Effective Dicke model}\label{Sec: the system}

\begin{figure}
\centering
\includegraphics[width=0.4\textwidth]{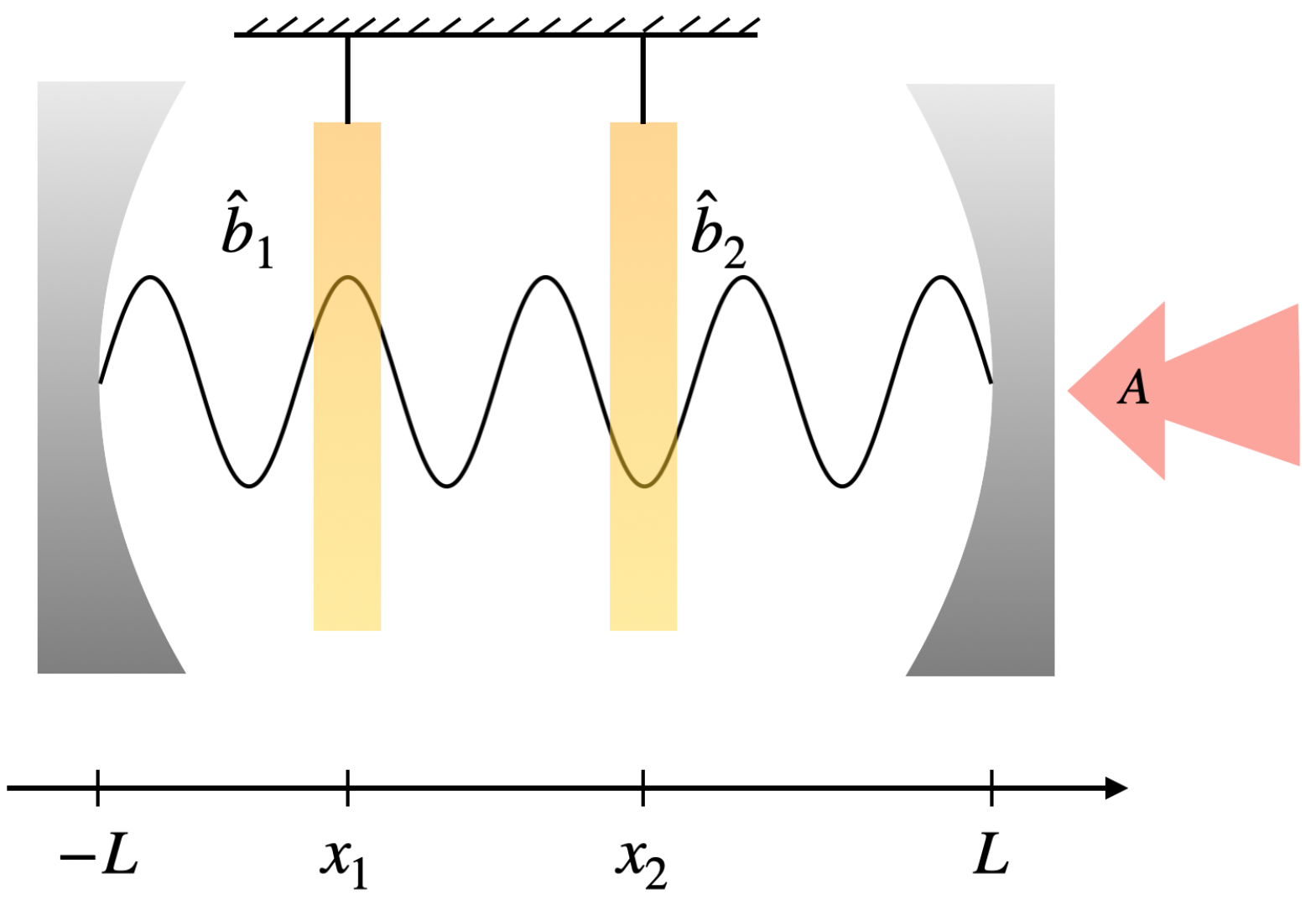}     
\caption{Schematics of the system, with two membranes inside a driven Fabry-Perot cavity. The positions $x_{1,2}$ of the mirrors are discussed in Appendix~\ref{Appendixa}.}  
\label{modelcp}
\end{figure}

The optomechanical system we consider is formed by two mechanical membranes inside a driven optical cavity, schematically shown in Fig.~\ref{modelcp}. The two membranes are located at antinodes of the cavity field, such that only second-order optomechanical couplings are significant. The Hamiltonian reads (setting $\hbar=1$):
\begin{equation}\label{H}
\hat H =\omega_{c}\hat a^{\dagger}\hat a+\omega_{1} \hat b_{1}^{\dagger}\hat b_{1}+\omega_{2}\hat b_{2}^{\dagger}\hat b_{2}+\hat H_I+\hat H_D,
\end{equation}
where $\omega_{1,2}$ ($\omega_{c}$) is the frequency of the relevant mechanical modes (cavity mode) with annihilation operators $\hat b_{1,2}$ ($\hat a$). The interaction ($\hat{H}_I$) and drive ($\hat{H}_D$) Hamiltonians are given by:
\begin{align}
\hat H_I =& g \hat a^{\dagger}\hat a(b_{1}^{\dagger}+\hat b_{1})^{2}-g\hat a^{\dagger}\hat a(\hat b_{2}+\hat b_{2}^{\dagger})^{2} \nonumber \\ 
&+J(\hat b_1^\dag +\hat b_1-\hat b_2^\dag -\hat b_2)^{2}, \label{HI} \\ 
\hat H_D =&A(\hat a e^{i\omega_D t}+\hat a^{\dagger}e^{-i\omega_D t}). \label{HD}
\end{align}
Here the second-order optomechanical couplings of the cavity with the two membranes are assumed to have opposite values ($g_1=-g_2=g$). The feasibility of this condition is discussed in detail in Appendix~\ref{Appendixa}), while the case $g_1\neq g_2$ will be considered in Sec.~\ref{Sec: time-crystal}. $J$ is the direct coupling between the two membranes which can be implemented through a 
coupling overhang~\cite{Spletzer2006_APL88-254102,Karabalin2009_PRB79-165309,Yabuno2013_APL103-063104,Okamoto2013_Nature9-480,Hao2018_PRApplied9-054024}. In $\hat H_D$, the parameters $\omega_D$ and $A$ are the frequency and the amplitude of the drive, where $A=\sqrt{2P_{L}\kappa/\omega_c}$ depends on the power of the drive $P_{L}$ and the decay rate of the cavity field $\kappa$. In a conventional Fabry-Perot cavity and for $50 \mathrm{pg}$ mechanical beams, one typically has $\omega_{i}/2\pi\sim100~\mathrm{kHz} ~ (i=1,2)$ and $g\sim \mathrm{Hz}\sim 10^{-6}\omega_i$~\cite{Thompson2008_Nature452-72,Sankey2010_NaturePhy6-707}. In addition, the structural coupling $J$ can reach up to $J\sim 10^{-2}\omega_{i}$~\cite{Karabalin2009_PRB79-165309,Okamoto2013_Nature9-480,Hao2018_PRApplied9-054024}, which  is much larger than a typical decay rate of the membranes, $\gamma/\omega_i \sim10^{-4}-10^{-7}$~\cite{Aspelmeyer2014_RevModPhys86-1391}. For the moment, we will neglect the small damping of the mechanical modes. Effects of a finite decay rate $\gamma$ will be discussed in Sec.~\ref{The membrane's decay is considered}.


We now show that, under appropriate conditions, the above optomechancial system becomes equivalent to the Dicke model, describing the interaction of a cavity mode with an ensamble of identical two-level systems. At strong drive, the optical cavity mode $\hat a$ can be decomposed ias $\hat{a}=(\alpha+\hat d')e^{-i\omega_D t}$, where $\hat d'$ represents the quantum fluctuations and $\alpha$ is the large classical amplitude of the driven cavity mode:
\begin{equation}\label{alpha}
\alpha =\frac{A}{i\kappa-\Delta}\equiv |\alpha|\exp[i\theta],
\end{equation} 
with $\Delta=\omega_{c}-\omega_D$ the detuning, which we choose positive. The interaction $\hat{H}_I$ leads to modified mechanical frequencies, $\tilde{\omega}_{1(2)}=\omega_{1(2)}+2J \pm 2g|\alpha|^{2}$. Considering a working point with equal effective frequencies:
\begin{equation}\label{constrain}
\omega_{1}+2J+2g|\alpha|^{2}=\omega_{2}+2J-2g|\alpha|^{2}\equiv\omega_{m},
\end{equation} 
we can derive the following effective Hamiltonian in the rotating frame $U=e^{-i\omega_D t \hat a^{\dagger}\hat a -i\omega_{m}t(\hat b_{1}^{\dagger}\hat b_{1}+\hat b_{2}^{\dagger}\hat b_{2})}$:
\begin{align}\label{effectiveH}
\hat H_{\rm{eff}}= & \Delta \hat d^{\dagger}\hat d +2g|\alpha|(\hat d+\hat d^{\dagger})(\hat b_{1}^{\dagger}\hat b_{1}-\hat b_{2}^{\dagger}\hat b_{2})\nonumber\\
 & -2J(\hat b_{1}\hat b_{2}^{\dagger}+\hat b_{1}^{\dagger}\hat b_{2}),
\end{align}
where we have defined $\hat{d} \equiv \hat d'\exp[-i\theta]$. Here, taking the rotating wave approximation (RWA), high frequency oscillating terms were neglected, leading to an effective Hamiltonian where the number of total phonons is conserved, i.e., $[\hat N,\hat H_{\rm{eff}}]=0$ ($\hat N=\hat b_1^\dag \hat b_1+\hat b_2^\dag \hat b_2$). We see that, applying the Schwinger's representation to Eq.~(\ref{effectiveH}), $\hat J_{z}=-(\hat b_{1}^{\dagger}\hat b_{2}+\hat b_{2}^{\dagger}\hat b_{1})/2,\hat J_{x}=(\hat b_{1}^{\dagger}\hat b_{1}-\hat b_{2}^{\dagger}\hat b_{2})/2$, the mechanical degrees of freedom can be written in terms of spin variables:
\begin{equation}\label{H_eff}
\hat H_{\rm{eff}}= \Delta \hat d^{\dagger}\hat d+4J\hat J_z+4g|\alpha|(\hat d+\hat d^{\dagger})\hat J_x.
\end{equation}
Thus, the system is mapped onto a Dicke model with a dissipative cavity. In a standard notation \cite{Dimer2007_PRA75-013804,Kirton2019_AdvancedQuantumTechnologies2-1970013}: 
\begin{equation}\label{H_DM}
H_{\rm{DM}}=\omega_0\hat c^\dagger c+\omega_z \hat J_z+\frac{2\lambda}{\sqrt{N_a}}(\hat c+\hat c^\dagger)\hat J_x,
\end{equation}
where $\hat c$ is the cavity mode and $\hat J_{z/x}=\frac12 \sum_{i=1}^{N_a} \hat \sigma_{z/x}^i$ are collective atomic operators, with $\hat\sigma_\alpha$ the Pauli matrices. The mapping yields $\omega_0 = \Delta$ and $\omega_z=4J$ for the cavity and atomic frequencies, respectively. The coupling strength is given by $\lambda = 2 g |\alpha|\sqrt{N_a}$, where the size $N_a$ of the atomic ensemble can be identified with the number $\hat{N}$ of mechanical excitations.

It is worth pointing out that another Dicke model realization in optomechanics has been proposed, considering a `membrane-in-the middle' setup~\cite{Mumford2015_AnnalenderPhysik527}. In that case, however, the roles of optical and mechanical degrees of freedom are switched, as the Dicke model cavity is mapped to a single mechanical membrane. Conversely, the spin ensemble is mapped to a pair of cavity modes. Therefore, in such realization the total angular momentum of the atomic ensemble decays to zero quickly, due the large damping of the optomechancal cavities. In contrast, in our system the role of the spin ensemble is played by the phonon modes, whose damping can be $10^5-10^6$ times smaller than $\kappa$~\cite{Aspelmeyer2014_RevModPhys86-1391,Thompson2008_Nature452-72}. Thus, the Dicke model is implemented in a more standard scenario. 

\begin{figure}
\centering
\includegraphics[width=0.4\textwidth]{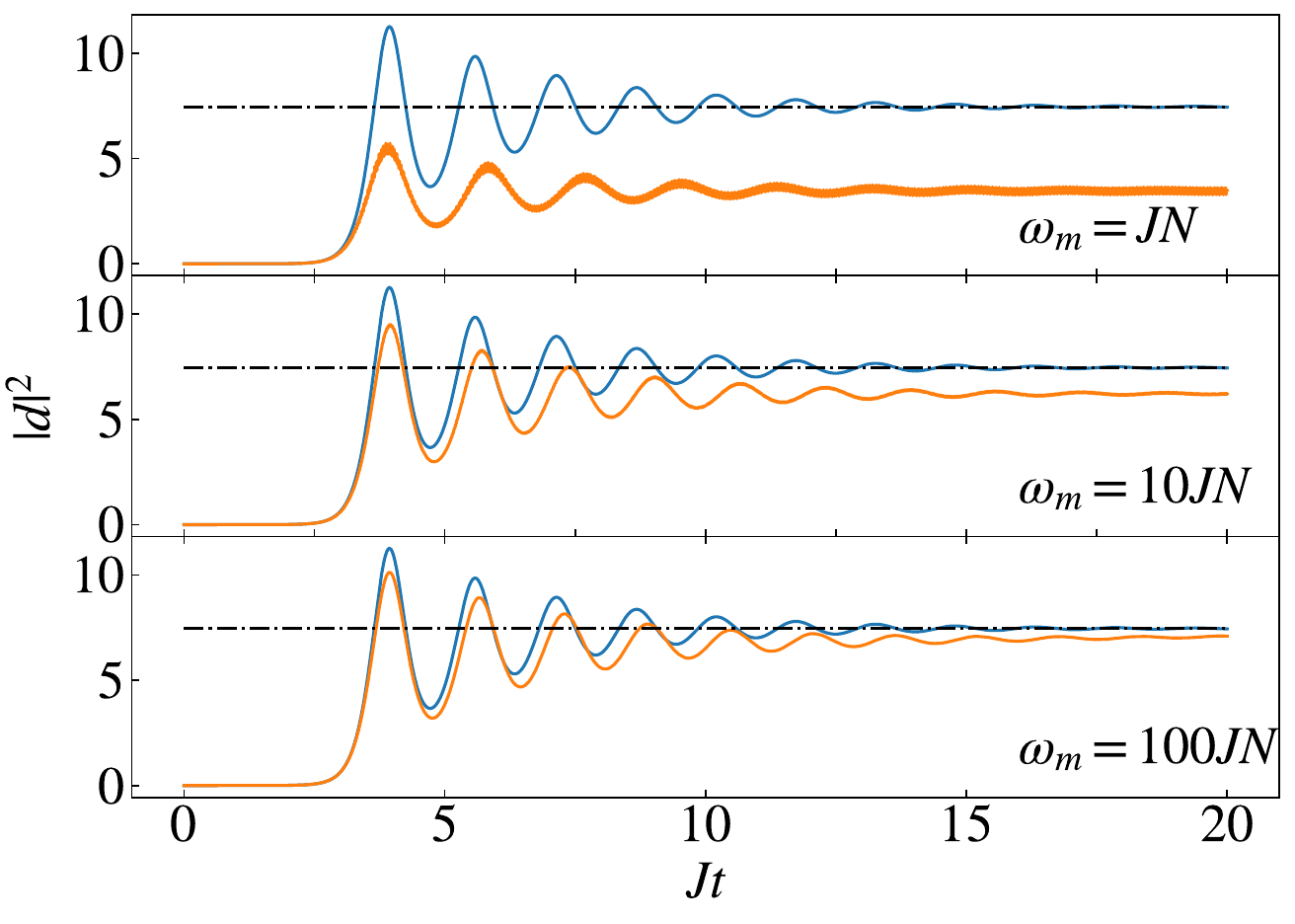}   
\caption{Validity of the effective model.  In each panel, the lower (orange) curve is obtained computing $|a-\alpha|^2$ from the mean-field equations of the full model and the upper (blue) curve is obtained from the effective Hamiltonian. The black dot-dashed lines refer to the mean-field steady-state of the effective model, given by Eq.~(\ref{orderp_d}). The three panels are computed for $N=200$ and $\omega_m/NJ = 1,10,100 $ (from top to bottom). Other parameters are: $A/J=2000$, $\Delta/J=20$, $\kappa/J=10$, $b_1(0)=b_2(0)=10$, and $g=1.2 g_c$. We compute $g_c$ as in Eq.~(\ref{criticalp}), while $\omega_1, \omega_2$ are decided by Eq.~(\ref{constrain}).}
\label{ana-num in new model}
\end{figure}

In the rest of the paper, we will discuss the quantum phase transition and a protocol to realize a discrete time crystal based on our setup with two membranes.  The validity of the effective model Eq.~(\ref{effectiveH}) can be tested through the mean-field approximation of the equations of motion:
\begin{eqnarray}\label{langevinequation}
i\frac{d}{dt}b_{1} & = & 2g|\alpha|(d+d^{*})b_{1}-2Jb_{2}\nonumber,\\
i\frac{d}{dt}b_{2} & = & -2g|\alpha|(d+d^{*})b_{2}-2Jb_{1}\nonumber,\\
i\frac{d}{dt}d & = & \Delta d+2g|\alpha|(b_{1}^{*}b_{1}-b_{2}^{*}b_{2})-i\kappa d,
\end{eqnarray}
where $\langle \hat b_1\rangle=b_1$, $\langle \hat b_2\rangle=b_2$, $\langle \hat d\rangle=d$. Here, quantum fluctuations are neglected and the factorization of expectation values is imposed, $\langle(\hat d+\hat d^\dagger)\hat b_{i}\rangle\rightarrow\langle(\hat d+\hat d^\dag)\rangle \langle \hat b_{i}\rangle$, $\langle\hat b_i^\dagger \hat b_i\rangle\rightarrow\langle \hat b_i^\dagger\rangle \langle \hat b_i\rangle$  (with $i=1,2$). Analogous equations can be derived from the full model, Eq.~(\ref{H}). A comparison between numerical results is shown in Fig.~\ref{ana-num in new model}, showing good agreement when $\omega_m$ is increased. This is because the two main approximations, linearization and RWA, require a sufficiently large $\alpha$ and $\omega_m\gg J(b_{i}^{2}+b_{i}^{\dagger2}),g|\alpha|^{2}(b_{i}^{2}+b_{i}^{\dagger2})$, respectively. 

\section{`Superradiant' phase}\label{Sec: Quantum phase transition}

\begin{figure}
\centering
\includegraphics[width=0.46\textwidth]{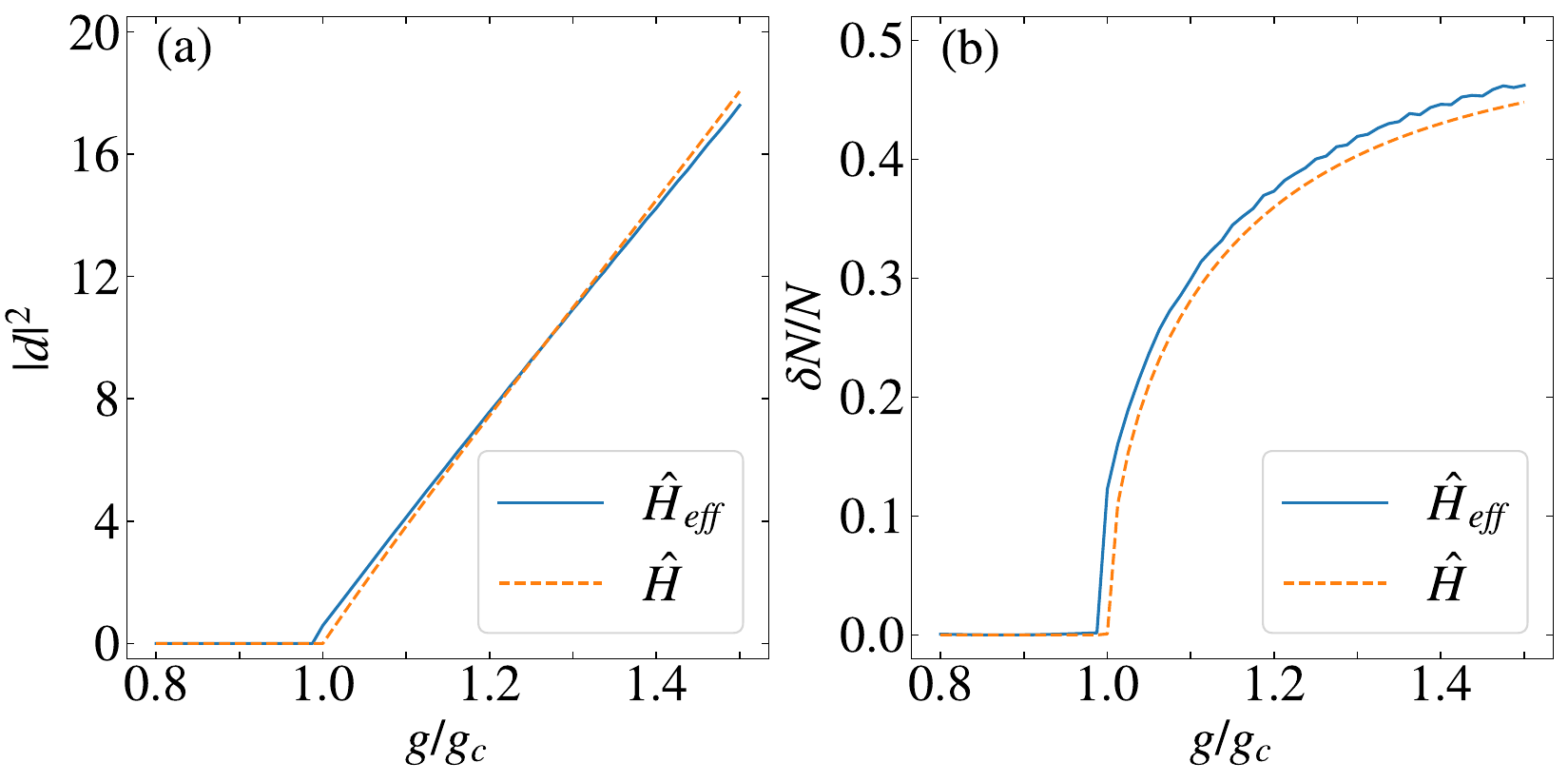}   
\caption{Dependence on $g$ of two order parameters. (a): Displacement $|d|^2$ of the cavity field. (b): Phonon difference between the two mechanical modes. In each panel, we compare stationary values from the effective Hamiltonian (blue lines) and simulations from the original Hamiltonian (orange dashed lines). We have used the following parameters: $\Delta/J=20$, $A/J=2\times 10^3$, $\kappa/J=10$, $b_1(0)=b_2(0)=10$, $\omega_m/J=10^4$. The final time of the simulations is $t_f =80/J$ and  $\omega_{1,2}$ are decided by Eq.~(\ref{constrain}).}
\label{PT in new model}
\end{figure}

In the thermodynamic limit $N_a \rightarrow \infty$, the Dicke model in Eq.~(\ref{H_DM}) displays a second-order phase transition from the normal phase to a superradiant phase, where the $\mathcal{Z} _2$ symmetry [defined by $\{\hat c,\hat \sigma_x\}\rightarrow\{-\hat c,-\hat \sigma_x\}$] is spontaneously broken, and both cavity field and atomic ensemble acquire macroscopic occupations. Such quantum phase transition has been demonstrated theoretically and experimentally~\cite{Dimer2007_PRA75-013804, Zhang2017_Optica4-424}. The analysis of the critical behavior can be performed using the mean field solution~\cite{Dimer2007_PRA75-013804, Kirton2017_PRL118-123602, Kirton2019_AdvancedQuantumTechnologies2-1970013}, which is valid in the thermodynamic limit, and gives the critical coupling~\cite{Emary2003_PRE67-066203}
\begin{equation}\label{criticalDicke}
\lambda_c=\sqrt{(\omega_0^2+\kappa^2)\omega_z/4\omega_0}.
\end{equation}
From the mapping detailed in the previous Section, we obtain the corresponding critical point of the optomechanical model:
\begin{equation}\label{criticalp}
g_{c}=\sqrt{\frac{(\Delta^{2}+\kappa^{2})J}{4|\alpha|^{2}N\Delta}},
\end{equation}
where $N\equiv |b_1|^2+|b_2|^2$ is the  (approximately) conserved total number of phonons. When the optomechancial coupling satisfies $g \leq g_c$, the system is in the normal phase with zero occupancy of the displaced cavity mode, $d=0$, and symmetric phonon numbers, $|b_1|^2=|b_2|^2$. For $g > g_c$, the $\mathcal{Z} _2$ symmetry is spontaneously broken and the system enters the superradiant phase, which implies a finite expectation value of $d$ and unbalanced phonon occupations, i.e., $\delta N =(|b_1|^2-|b_2|^2)/2\neq 0$. The two symmetry-broken states are $|\alpha \mp \bar{d}, \pm \delta \bar{N}\rangle$, with the stationary values 
\begin{eqnarray}
\bar{d}&=&\frac{2g|\alpha|N}{\Delta-i\kappa}\sqrt{1-\frac{g_c^4}{g^4}},  \label{orderp_d}\\
\delta \bar{N} &=&\frac{N}{2}\sqrt{1-\frac{g_c^4}{g^4}}.   \label{orderp_dN}
\end{eqnarray}
As shown by the dashed lines in Fig.~\ref{ana-num in new model}, the finite expectation value of Eq.~(\ref{orderp_d}) is exact for the effective model and shows good agreement with the stationary value of the full model, in the expected regime of validity. A more detailed comparison of order parameters across the critical coupling is presented in Fig.~\ref{PT in new model}, showing good agreement between analytical expressions and simulations from the original Hamiltonian. 

To obtain the above results directly from the mean-field equations~(\ref{langevinequation}), without invoking the mapping to the Dicke model, one should consider the ansatz $b_{1,2}\rightarrow \beta_{1,2}e^{i\omega t}$. Here $\beta_{1,2}$ are stationary amplitudes and the effective mechanical frequency is given by:
\begin{equation}\label{effective_omega}
\omega=2J \left[ 1+\frac{g^2-g_c^2}{g_c^2}\theta(g-g_c)  \right]. 
\end{equation}
Equation~(\ref{effective_omega}) reflects the fact that, for $g\leq g_c$ the two membranes are coupled oscillators with a symmetric normal mode of frequency $\omega=2J$ (in the rotating frame). However, the coupling to the cavity should be taken into consideration in the superradiant phase, which modifies the effective frequency of the normal mode. We refer to Appendix~\ref{Appendixb} for explicit calculations.

Finally, we comment on the role of mechanical dissipation. If the decay of the membranes is considered, the total phonon number is not a conserved quantity but slowly decays with time. Supposing to start from the superradiant phase, and keeping the strength and detuning of the external drive fixed, one finds that the critical coupling $g_c$ slowly grows in time, due to the decrease of $N$ [see Eq.~(\ref{criticalp})]. During this slow evolution, thanks to the large cavity damping, the system follows adiabatically the broken-symmetry state. Correspondingly, the order parameters of Eqs.~(\ref{orderp_d}) and (\ref{orderp_dN}) gradually decrease. Finally, when the critical point $g_c$ becomes larger than the fixed coupling strength $g$, the system recovers the normal phase. From this qualitative description we see that a finite mechanical damping allows in principle to observe the phase transition in time domain. As the same behavior occurs for the time-crystal, we defer a more detailed discussion to Sec.~\ref{The membrane's decay is considered}. See, in particular, Fig.~\ref{DTC-gammadecay}.

\section{Discrete time crystal}\label{Sec: time-crystal}

The realization of a time crystal in cavity/circuit QED systems, based on a Dicke model with tunable coupling, has been recently proposed in Ref.~\cite{Ueda2018_PRL120-040404}. The basic idea is to periodically control the dipole interaction and alternate finite coupling and free evolution periods. In an ideal limit, assuming small dissipation and the resonant condition $\omega_z=\omega_0$, the system is in one of two stationary broken-symmetry states for $\lambda>\omega_0/2$. Subsequently, free evolution for a period $\pi/\omega_0$ accumulates a $\pi$ phase, which switches the system from one steady-state to the other. The repetition of this protocol in time generates a discrete  dissipative time crystal, robust to deviations from the ideal limit \cite{Ueda2018_PRL120-040404}. 

However, this idea cannot be applied in a straightforward manner to our optomechancal system. Setting $\lambda=0$ in Eq.~(\ref{H_DM}) corresponds $\alpha=0$, since the bare optomechanical coupling $g$ in Eq.~(\ref{H_eff}) is not easily modified. But turning off the external drive invalidates the resonant condition Eq.~(\ref{constrain}), on which the mapping from $\hat{H}$ to $\hat{H}_{\rm eff}$ is based. To circumvent this problem, we notice that a free evolution is not necessary, as an equivalent result can be achieved by tuning parameters to the normal phase. Even if the ensuing dynamics is more complex, due to nonlinear features of the (still interacting) model, an approximate $\pi$ rotation can be realized in this manner. Such incomplete flip is sufficient to establish a discrete time crystal, due to its intrinsic robustness to imperfections. In practice, we consider below a protocol where the critical point $g_c$ is modified through a simultaneous adjustments of detuning $\Delta$ and drive amplitude $A$ [cf.~Eq. (\ref{criticalp})]. This allows us to drive the optomechanical system to the normal phase by keeping $g$ and $\alpha$ fixed, thus preserving the mapping to the Dicke model.

\subsection{Period-doubling Floquet dynamics}\label{Sec: realizationprocess}

\begin{figure}
\centering
\includegraphics[width=0.46\textwidth]{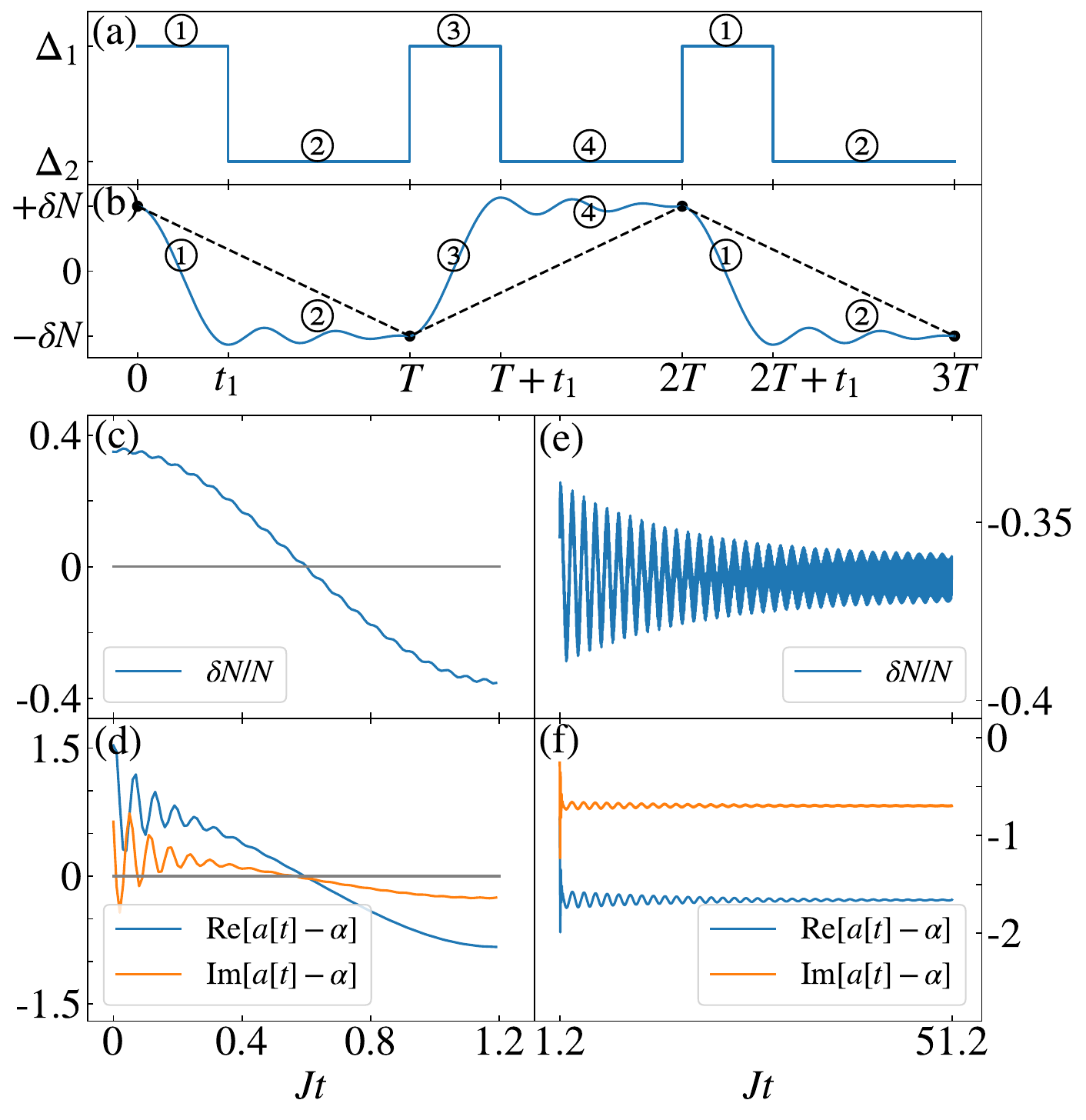}
\caption{Discrete time-crystal protocol. The time dependence of detuning, with the corresponding time evolution of $\delta N$, are illustrated in panels (a) and (b), respectively. The blue line in panel (b) depicts the continuous-time evolution of the phonon difference and the black dashed line is the stroboscopic dynamics at $t=kT\ (k=0,1,2,\cdots)$. Panels (c) and (d) show the actual time evolution of $\delta N$ and $d$ during the flipping process \textcircled{1}, obtained from numerical simulations. Panels (e) and (f) show the time evolution during the relaxation process \textcircled{2}.  We have used: $\Delta_1/J=100$, $\Delta_2/J=50$, $A_2/J=10^4$, $\kappa/J=10$, $N=200$, $g=1.2g_{c,2}$, $\omega_m/J=10^4$, $t_1=1.196/J$ and $t_2=100/J$, while $A_1$ is given by Eq.~(\ref{A2func}).}
\label{DTCprocess}
\end{figure}

In Sec.~\ref{Sec: Quantum phase transition}, we have discussed how the model with two membranes in the cavity exhibits a second-order phase transition in the thermodynamical limit of infinite phonon number, $N\rightarrow\infty$, when $g>g_{c}$. Based on this phase transition, an exact period-doubling Floquet dynamics can be constructed with 4 basic steps, illustrated in Fig.~\ref{DTCprocess}(a). After initializing the system in one of the two symmetry-broken steady-states, say  $|\alpha-\bar{d}, \delta \bar{N}\rangle$, the protocol reads as follows: 

\textcircled{1}  Detuning and drive amplitude are switched to $\Delta_1$ and $A_1$, respectively, such that $g< g_{c,1}$ and the system is in the normal phase. Importantly, the new values should satisfy $\alpha = A_1/(i\kappa-\Delta_1)$ (i.e., the amplitude of the initialization step remains unchanged). Now the system undergoes an oscillatory dynamics, shown in panels (c) and (d) of Fig.~\ref{DTCprocess}. For a proper evolution time $t_1$ (the choice of $t_1$ will be discussed in detail in Sec.~\ref{sec:flipping_time}), the state is approximately flipped from $|\alpha-\bar{d}, \delta \bar{N}\rangle$ to $|\alpha+\bar{d}, -\delta \bar{N}\rangle$. We note that the effective oscillator should not be in the overdamped regime, otherwise it will simply relax to $|\alpha,\delta N=0 \rangle$. Even if an oscillatory dynamics takes place, an exact flip is usually not possible.

\textcircled{2} The detuning is changed to $\Delta_2$ and the drive amplitude to $A_2$, still satisfying
\begin{equation}
\label{A2func}
\frac{A_2}{i\kappa-\Delta_2}=\frac{A_1}{i\kappa-\Delta_1}.
\end{equation}
The values of $\Delta_2$ and $A_2$ can simply be the initialization values. They are chosen to yield $g > g_{c,2}$ but do not affect the classical amplitude $\alpha$ of the cavity, see Eq.~(\ref{alpha}). Now the system is in superradiant phase for a time $t_2$, during which it relaxes towards the nearby steady-state, i.e., $|\alpha+\bar{d}, -\delta \bar{N}\rangle$. 

\textcircled{3} Setting $\Delta=\Delta_1$ and $A=A_1$ for a time duration $t_1$ returns the system to the normal phase, which induces an approximate evolution from $|\alpha+\bar{d}, -\delta \bar{N}\rangle$ to $|\alpha-\bar{d}, \delta \bar{N}\rangle$.

\textcircled{4} $\Delta=\Delta_2$ and $A=A_2$ for a time duration $t_2$. At the end of this step, the state is relaxed towards the initial state $|\alpha-\bar{d}, \delta \bar{N}\rangle$.

\begin{figure}
\centering
\includegraphics[width=0.46\textwidth]{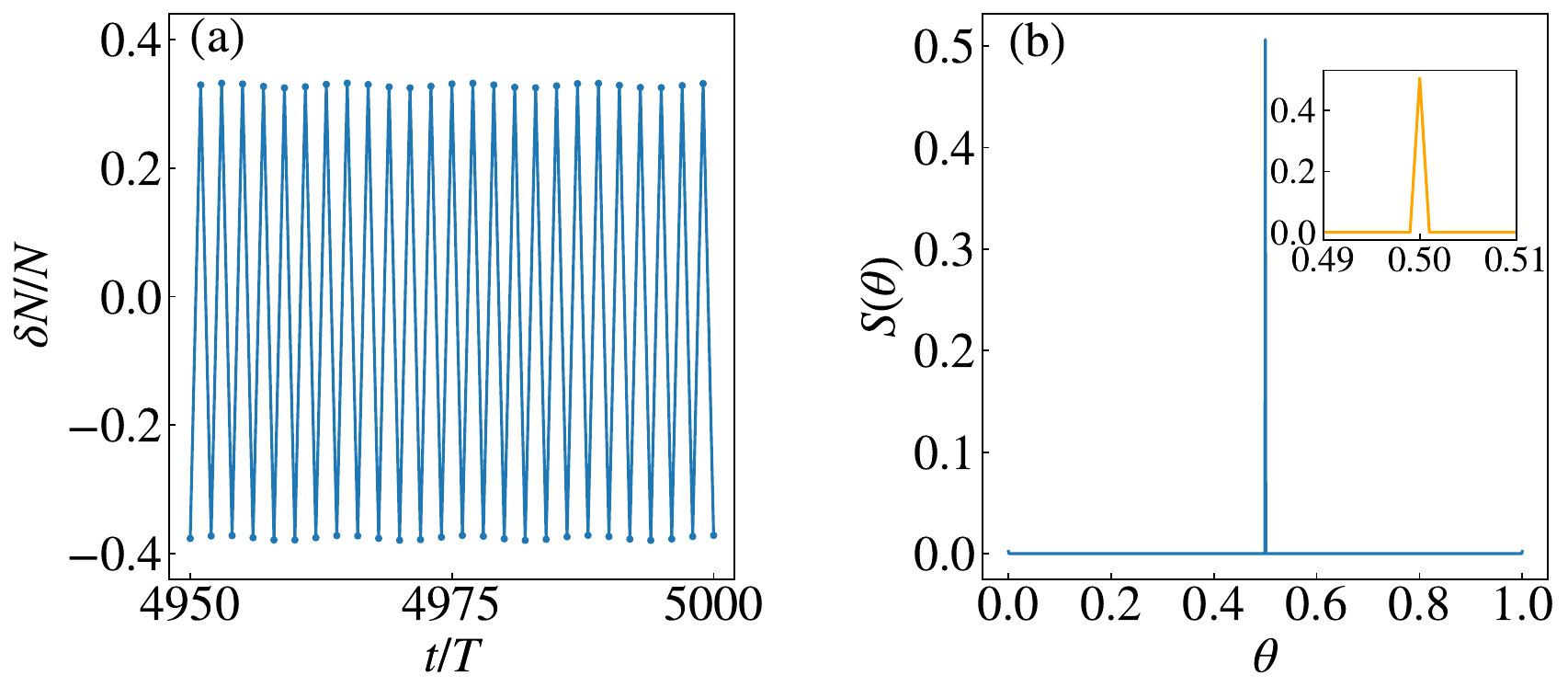}
\caption{Persistence of the stroboscopic dynamics. In panel (a) we show the stroboscopic time evolution of $\delta N/N$ in the interval $[4950,5000]\times T$. Panel (b) is the corresponding Fourier transform, given by Eq.~(\ref{FTdef}) with $n=5000$. The inset of (b) is a zoom-in of the peak. We used: $\kappa/J=1$, $\Delta_1/J=20$, $\Delta_2/J=5$, $A_2/J=600$, $g=1.2 g_{c,2}$, $\omega_m/J=3 \times 10^3$, $N=200$, $t_1=1.22/J$ and $t_2=30/J$.}
\label{longtimeDTC}
\end{figure}

In summary, the periodic change in detuning and drive amplitude is described by
\begin{align}
\label{DAfunc}
\{\Delta,A\}= & \begin{cases}
\{\Delta_{1},A_1\} & \left[0,t_{1}\right),\\
\{\Delta_{2},A_2\} & \left[t_{1},T\right),
\end{cases}
\end{align}
where the two pairs are related as in Eq.~(\ref{A2func}). The system returns to the initial state with period $2T$ (where $T=t_1+t_2$), thus doubling the period of the control pulse. An example of persistent period-doubling behavior induced by the above control pulse is shown in Fig.~\ref{longtimeDTC}, through the stroboscopic dynamics of $\delta N/N$ and its discrete Fourier transform: 
\begin{equation}\label{FTdef}
S(\theta)=\frac{1}{n}\sum_{k=1}^n \frac{\delta N(k)}{N}\exp(i 2\pi k\theta),
\end{equation}
defined as in~Refs.~\cite{LaRocca2020_PRA101-013417, Yao2017_PRL118-030401}. Here, $\delta N(k)$ is the phonon difference between the two membranes at the end of k-th period. Note that the stroboscopic oscillation in Fig.~\ref{longtimeDTC}(a) is not strictly symmetric around $\delta N=0$. As the original Hamiltonian is not exactly ${\mathcal{Z}}_2$-symmetric, the asymmetry reflects small corrections to the effective Dicke model (\ref{H_eff}).

\begin{figure}
 \centering
\includegraphics[width=0.4\textwidth]{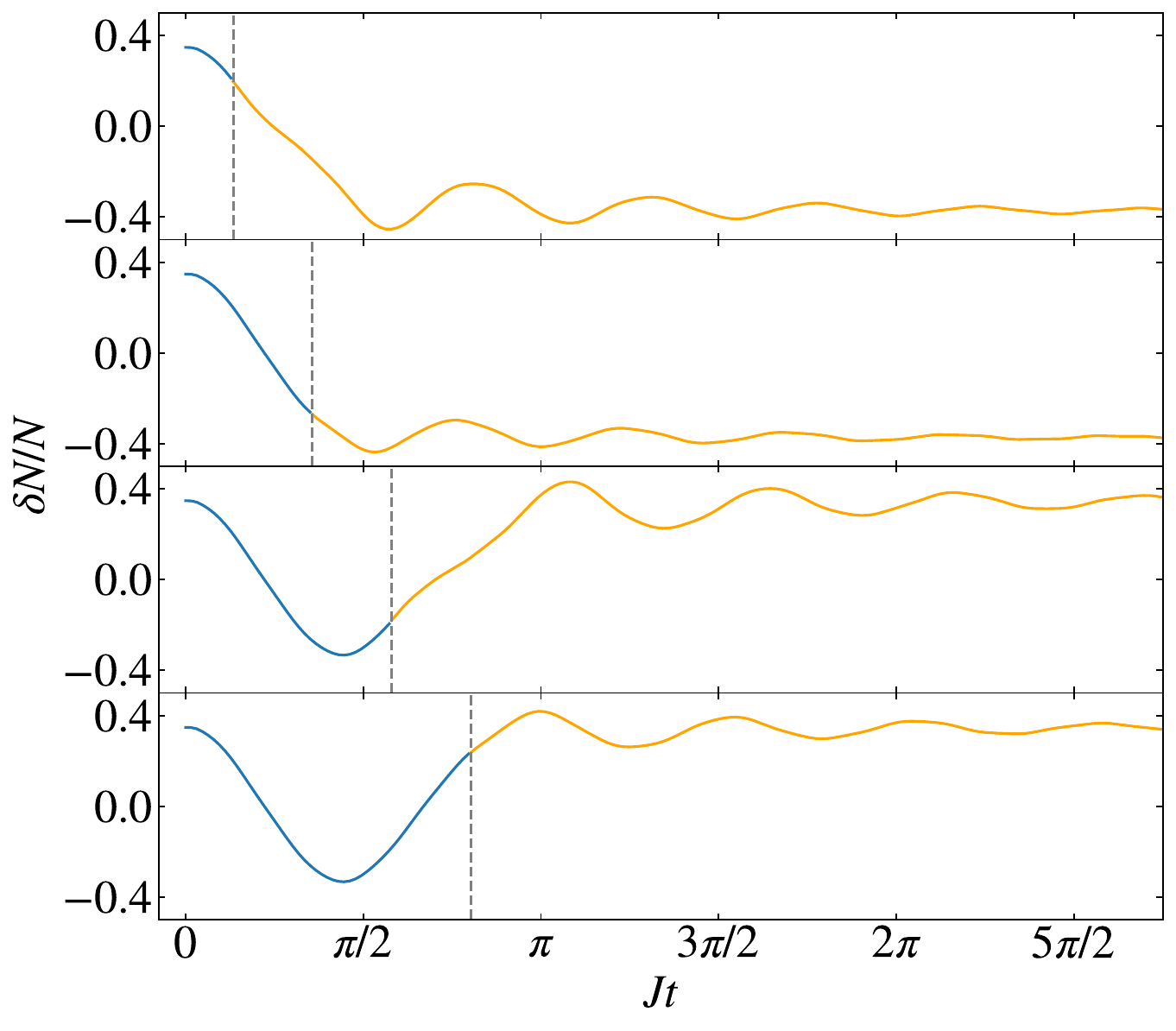}
\caption{Time evolution for different choices of the flipping time $t_1$. The gray dashed lines mark $Jt_1=\{0.7,1.4,2.1,2.8\}$ (from top to bottom). The time evolution at $t>t_1$ (orange curves) is in the relaxation phase \textcircled{2}. We used: $\kappa/J=10$, $\Delta_1/J=40$, $\Delta_2/J=20$, $A_2/J=2\times 10^3$, $N=200$, $g=1.2 g_{c,2}$, and $\omega_m/J=10^4$.}
\label{precession}
\end{figure}

\subsection{The choice of flipping time}\label{sec:flipping_time}
\begin{figure}
\centering
\includegraphics[width=0.4\textwidth]{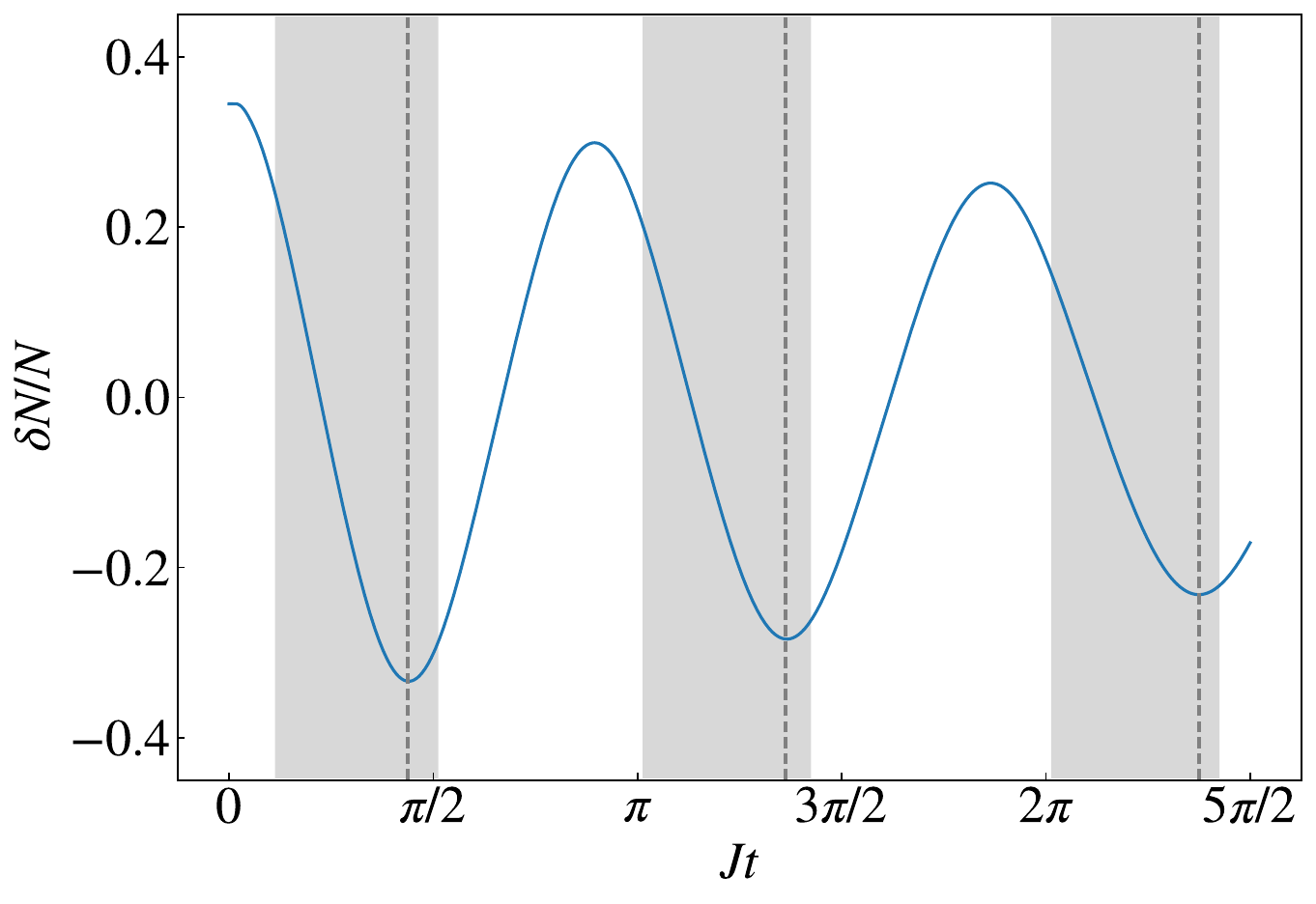}  
\caption{Flipping times leading to DTC dynamics. The blue line represents the time evolution during the flipping process \textcircled{1} (without the restriction $t<t_1$). DTC behavior can be obtained by choosing the flipping time $t_1$ inside the shaded regions. Dashed lines mark the minima of $\delta N/N$, around which the DTC regions are (asymmetrically) located. We used: $\kappa/J=10$, $\Delta_1/J=40$, $\Delta_2/J=20$, $A_2/J=2 \times 10^3$, $N=200$, $g=1.2 g_{c,2}$, $\omega_m/J=10^4$ and $J t_2$=50.}
\label{rigidity} 
\end{figure}

As explained above, each $2T$ operation cycle involves two flipping operations (\textcircled{1} and \textcircled{3}), where the steady-state $|\alpha\mp \bar{d}, \pm \delta \bar{N}\rangle$ flips to the other steady-state $|\alpha \pm \bar{d},\mp\delta \bar{N}\rangle$. In the presence of decoherence and the always-on interaction $g$, an analytical expression of the ideal flipping time is not readily available. However, suitable values of $t_1$ can be found numerically. We find that the choice of the flipping time is rather flexible, because the actual flipping operation has a certain inertia and continues into the relaxation process even after the system is driven back to the superradiant-phase parameters. To demonstrate this, the evolution in the flipping process \textcircled{1} (blue lines) is shown in Fig.~\ref{precession} for different choices of the flipping times $t_1$. We also continue the time evolution beyond $t_1$, into the relaxation process \textcircled{2} (orange lines).  One can see that the initial evolution in the relaxation processes \textcircled{2} is a continuation of the oscillatory dependence of \textcircled{1}. Since an imperfect flipping can be completed during the relaxation phase, see in particular the top panel of Fig.~\ref{precession}, the flipping time $t_1$ can be chosen in a wide range. In Fig.~{\ref{rigidity}} we mark by a shadowed region (we refer to this as “DTC region”) the ranges of $t_1$ which allow the persistent oscillatory behavior of a time crystal. From the third panel of Fig.~\ref{precession} we can also see that the continuation of the oscillatory dependence into the relaxation phase can bring back a nearly-flipped state to the starting point.  For this reason, the DTC regions of Fig.~{\ref{rigidity}}  appear at `advanced' times, instead of being centered at the minima of the time evolution (dashed lines). 

\subsection{Rigidity of the DTC}
In the previous subsection we have shown that, for the proposed control pulse, the optomechanical system exhibits long time oscillations with period doubling. To qualify as time crystal, this persistent oscillation must be robust against parameter deviations, i.e., it should not occur at a finely-tuned point in parameter space. This property is also important for the experimental realization, where imperfections are unavoidable. In this section, we discuss how the DTC phase is affected by variations of different parameters, such as detuning and optomechanical coupling. 

\begin{figure}
\centering
\includegraphics[width=0.4\textwidth]{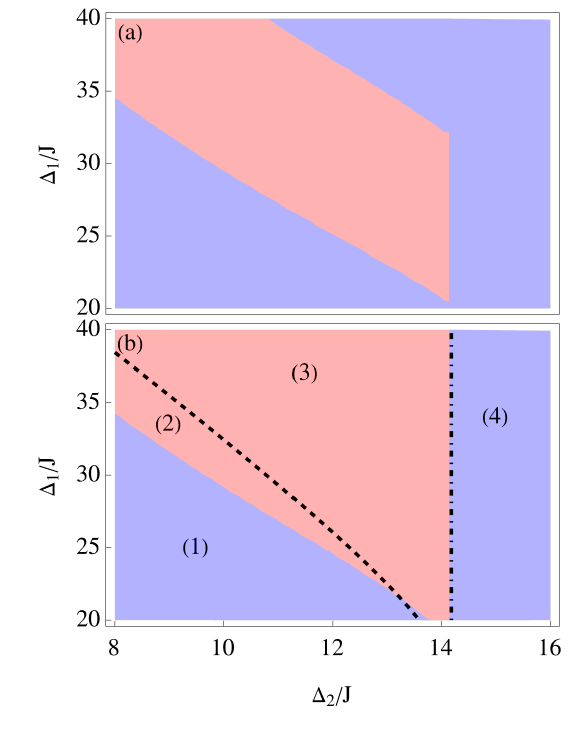}  
\caption{Phase diagram of DTC order with respect to $\Delta_1$ and $\Delta_2$. The red (blue) region refers to the DTC (non-DTC) phase. The two black lines indicate the conditions $g=g_{c,1}$ (dashed) and $g=g_{c,2}$ (dot-dashed). In panel (a) the flipping time is fixed at $t_1=1.90/J$, while in panel (b) $t_1$ is chosen for each $\Delta_{1/2}$ by the position of the first minimum in the time evolution of $\delta N$ (see, e.g., Fig.~\ref{rigidity}). Other parameters are the same for both panels: $\kappa/J=15$, $A_2/J=2000$, $g/J=2\times 10^{-3}$, $N=200$, $\omega_m/J=10^4$, $J t_2=200$.}
\label{phasediag1} 
\end{figure}

\begin{figure}
\centering
\includegraphics[width=0.35\textwidth]{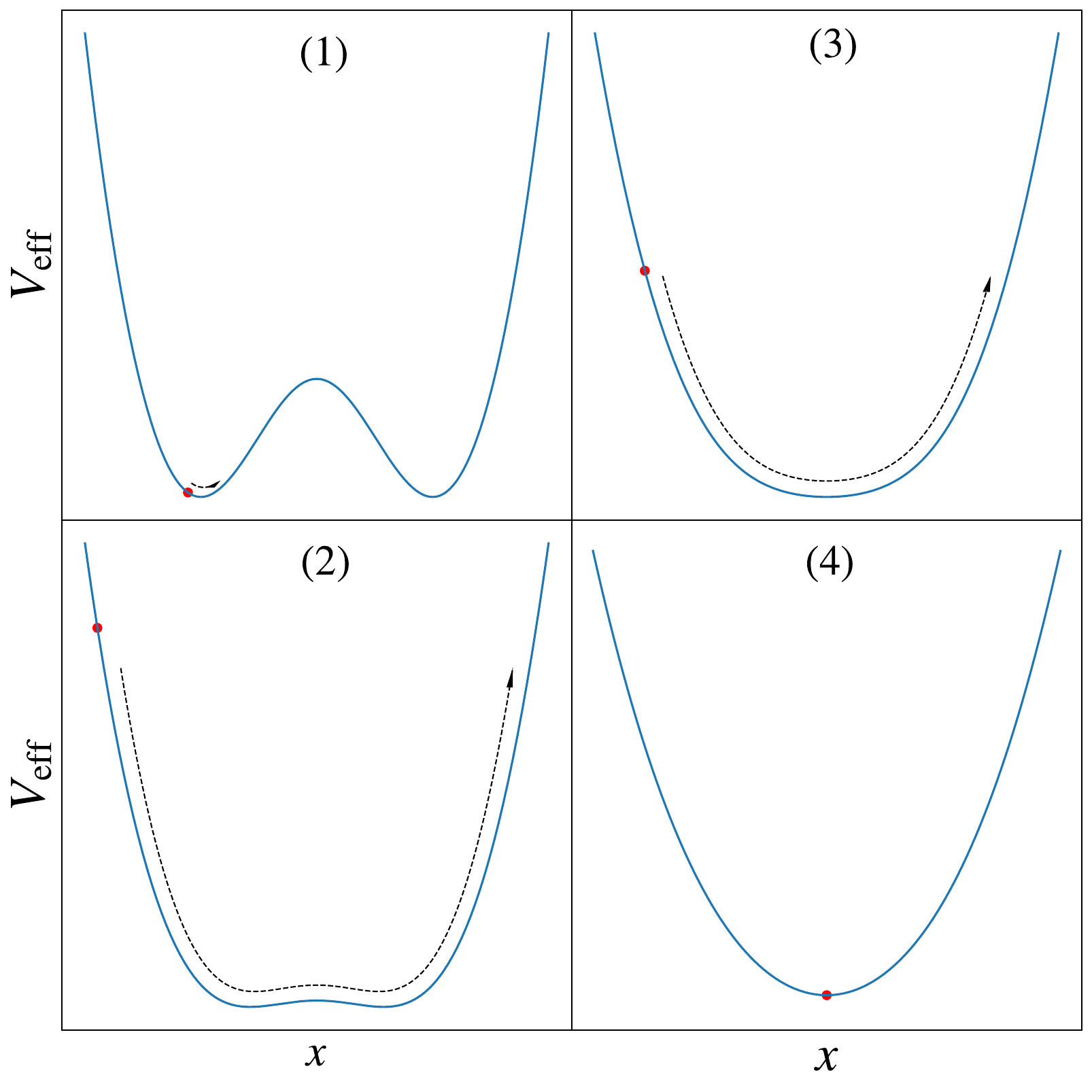}  
\caption{Schematic plot of the cavity effective potential $V_{\rm eff}(x)$ [cf. Eq.~(\ref{Veff})] during the flipping process, for four possible scenarios. The initial value of $x$ (indicated by a red dot) is decided by the previous relaxation process. Panels $(1)$-$(4)$ correspond to the four areas in the phase diagram of Fig.~\ref{phasediag1}.  }
\label{phasediag2} 
\end{figure}

Varying $\Delta_1$ and $\Delta_2$, we obtain the phase diagram of DTC order shown in Fig.~\ref{phasediag1}(a). Period-doubling occurs in the region marked in red, while in the blue region such behavior is absent. It is evident that period-doubling is robust to the imperfection in $\Delta_1$ and $\Delta_2$. In Fig.~\ref{phasediag1}(a) the flipping time $t_1$ is fixed, but a larger DTC region can be obtained if $t_1$ is further optimized at each point of the phase diagram, see Fig.~\ref{phasediag1}(b). Here the two black lines indicate the conditions $g_{c,1}=g$ and $g_{c,2}=g$. As explained above, when $g_{c,2}<g<g_{c,1}$ is satisfied, i.e., in region $(3)$, the DTC can be realized. However, DTC behavior also occurs in region $(2)$, where $g>g_{c,1}$. To understand the persistence of the DTC phase in region $(2)$ one can introduce an effective potential $V_{\rm eff}(x)$, where $x=(d+d^*)/\sqrt{2}$ is a quadrature of the cavity  (see Appendix~\ref{Appendixd} for the derivation). Various profiles of $V_{\rm eff}(x)$ during the flipping process are shown in Fig.~{\ref{phasediag2}}, where the red point is the initial position of the cavity and is decided by the previous relaxation process. The effective potentials in $(1)$-$(4)$ correspond to the four regions of Fig.~\ref{phasediag1}(b). The bottom left panel shows that in region $(2)$, although $g>g_{c,1}$ implies an effective potential with a double-well dependence, the barrier at $x=0$ is smaller than the initial value of the potential energy, thus does not prevent the flipping process from one steady-state $|\alpha-\bar{d},+\delta \bar{N}\rangle$ to the other steady-state $|\alpha+\bar{d},-\delta \bar{N}\rangle$ to take place.

\begin{figure}
\centering
\includegraphics[width=0.35\textwidth]{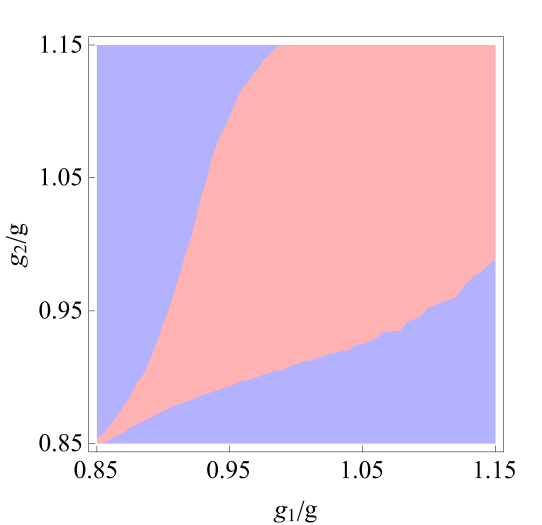}  
\caption{Phase diagram of DTC order with respect to $g_{1}$ and $g_2$. The red region is the DTC phase. We used: $\kappa/J=10$, $\Delta_1/J=100$, $\Delta_2/J=50$, $A_2/J=10^4$, $N=200$, $\omega_m/J=10^4$, and $Jt_2=100$. The two optomechanical couplings $g_{1,2}$ are expressed here in terms of $g=1.2g_{c,2}$, and $t_1$ is determined at each point of the phase diagram from the first minimum in the oscillations of $\delta N$ (see, e.g., Fig.~\ref{rigidity}).}
\label{phasediad-twog}
\end{figure}

Besides being robust to changes in $\Delta_1$ and $\Delta_2$, the period doubling oscillation is also tolerant to deviations from $g_1=g_2=g$ (i.e., the condition of equal optomechanical couplings), which is is particularly important for the experimental realization. A phase diagram of DTC order with respect to independent optomechanical couplings $g_{1,2}$ is shown in Fig.~\ref{phasediad-twog}. 

\subsection{DTC behavior in the deep quantum regime}

\begin{figure}
\centering
\includegraphics[width=0.46\textwidth]{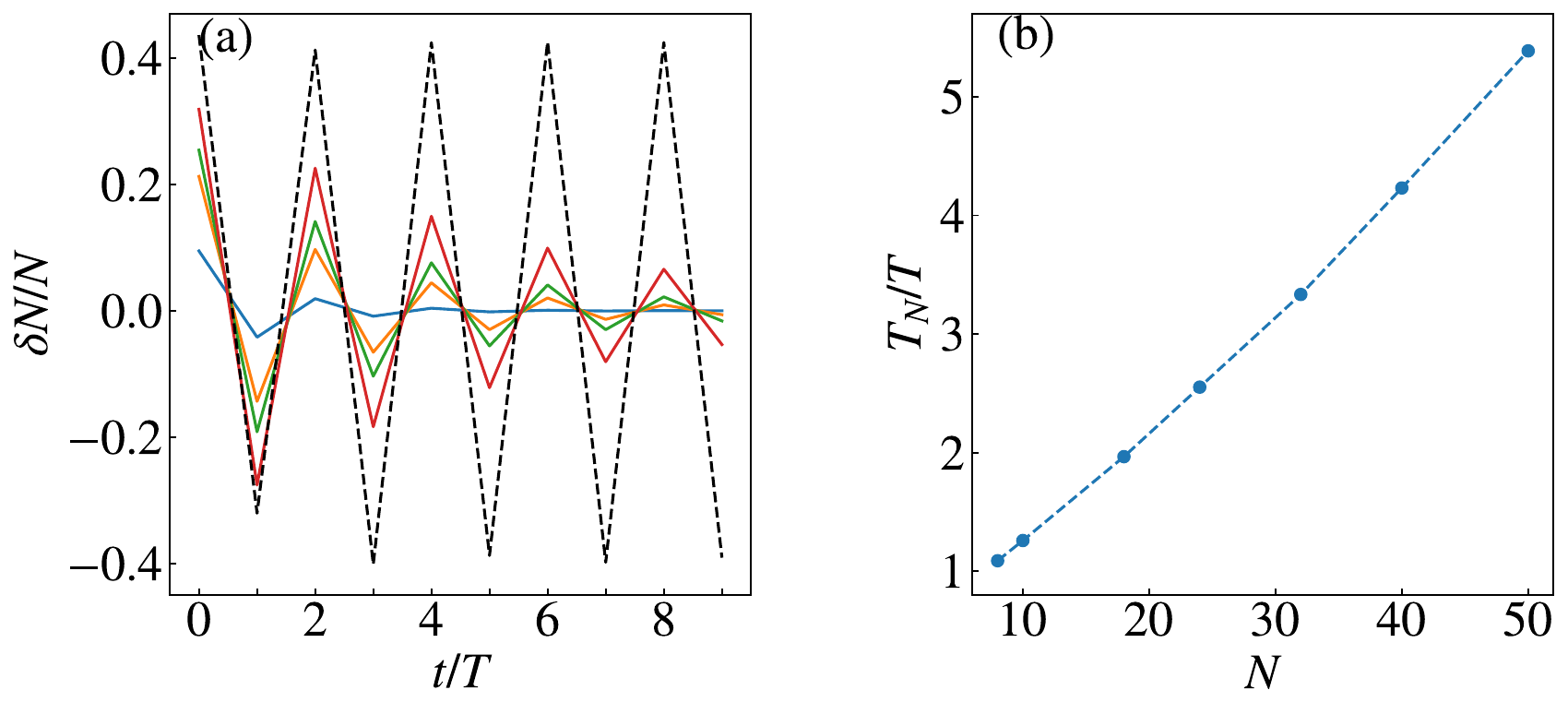}
\caption{Period-doubling oscillations in the deep quantum regime. In (a) we show the time evolution of $\delta N/N$ at different values of the (conserved) phonon number $N$. The purple line is the simulation from the original Hamiltonian, obtained by the mean-field approximation.  The solid lines are quantum simulations obtained from the master equation Eq.~(\ref{master_equation}) using $N=10,24,32,50$. A larger value of $N$ results in a larger oscillation amplitude. In panel (b) we plot the lifetime $T_N$ (dots), extracted from the simulations of panel (a). The dashed line is a guide for the eye. We used the following parameters: $\kappa/J=1.2$,  $\Delta_1/J=20$, $\Delta_2/J=5$, $A_2/J=300$, $\omega_m/J=1500$, $g=1.5 g_{c,2}$, $J t_1=5.94$, and $J t_2=5$. }
\label{smalnDTC}
\end{figure}

While time-crystal order appears in the thermodynamic limit, $N\rightarrow \infty$, experimental realizations are certainly limited to a finite excitation number. In principle, at finite $N$ the mean field approximation is not exact and numerical simulations with the full quantum treatment should be performed. In this regime of finite $N$, the period-doubling oscillation is only transient. However, their decay time diverges when increasing $N$. To address these effects, we investigate the few-phonon regime by solving the quantum master equation directly:
\begin{equation}\label{master_equation}
\frac{d\hat\rho}{dt}=-i[\hat H_{\mathrm{eff}},\hat \rho]+\kappa (\hat d\hat \rho\hat d^\dag-\frac{1}{2}(\hat d^\dag \hat d\hat \rho+\hat \rho\hat d^\dag \hat d)).
\end{equation}
For simplicity, we only perform simulation based on $\hat H_{\mathrm{eff}}$, expressed as in Eq.~(\ref{effectiveH}) through the Schwinger's representation~\cite{Johansson2012_ComPC183-1760}. Oscillations of $\langle\delta \hat N/N\rangle=\langle (\hat b_1^\dagger \hat b_1-\hat b_2^\dagger \hat b_2)/2N\rangle$ are displayed in panel (a) of Fig.~{\ref{smalnDTC}}, for different values of the (conserved) phonon number $N$. As expected, the oscillation period is $2T$ while the amplitude at given $N$ follows an approximate exponential decay $\sim e^{- t/T_{N}}$. By increasing $N$, we observe both a general increase of amplitude, bringing the oscillations closer to the mean-field result (dashed curve), as well as a longer decay time $T_N$. 

The growth of $T_N$ with $N$, shown in Fig.~{\ref{smalnDTC}}(b), is consistent with a robust DTC order in the thermodynamic limit. The dependence is slightly faster than linear in the available range of $N$, but the precise functional form is difficult to ascertain. If larger values of $N$ were accessible, $T_N$ might show the same type of weak exponential growth discussed in Ref.~\cite{Ueda2018_PRL120-040404}.  While it is numerically difficult to extend the simulations to larger $N$, we note that the total number of phonons is naturally large in our optomechanical model, which validates the thermodynamic limit and justifies the mean-field description adopted in the rest of the article. Instead, in Ref.~\cite{Ueda2018_PRL120-040404} the number of artificial atoms is typically of order $O(1)$.

 
\subsection{Mechanical dissipation}\label{The membrane's decay is considered}
\begin{figure}
\centering
\includegraphics[width=0.46\textwidth]{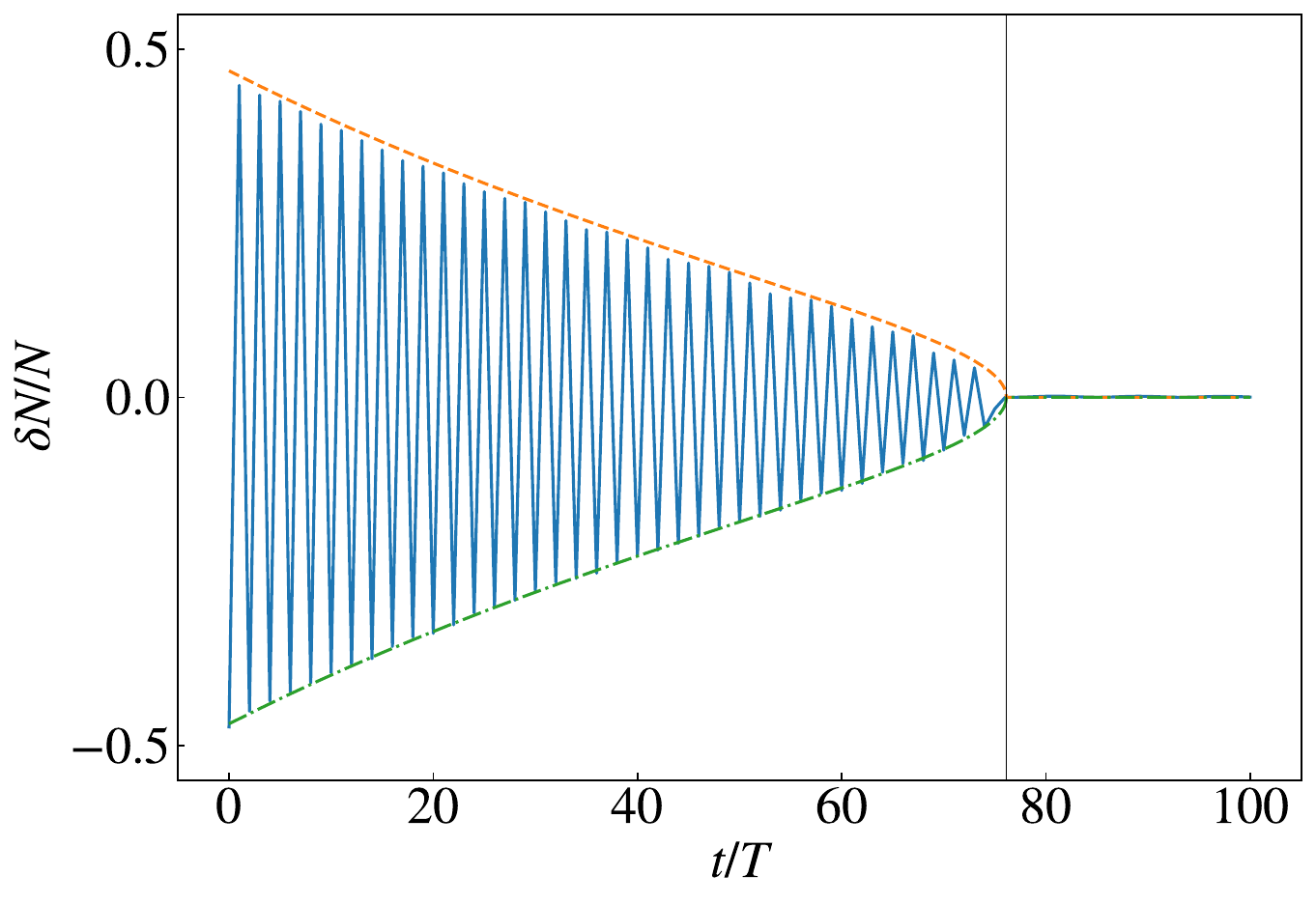}  
\caption{Evolution of the DTC order parameter $\delta N/N$ after including the decay of the mechanical modes. The blue curve is the numerical evolution, obtained from the mean-field equations of the full model, see Eq.~(\ref{modelcp}). The dashed and dot-dashed curves are from Eq.~(\ref{dnt}). We have used the following parameters: $\Delta_1/J=80$, $\Delta_2/J=20$, $A_2/J=2000$, $\kappa/J=10$, $\gamma/J=10^{-3}$, $\omega_m/J=10^4$, $Jt_1=0.97$, $Jt_2=6$, and $g=1.7g_{c,2}(0)$, where the critical coupling $g_{c,2}(0)$ is computed from Eq.~(\ref{criticalp}) using $N=200$ (the initial number of phonons). The value of $A_1$ is determined by Eq.~(\ref{A2func}).}
\label{DTC-gammadecay} 
\end{figure}

So far all our discussions have assumed negligible mechanical dissipation. Then, within the regime of validity of the effective Dicke model Eq.~(\ref{H_eff}), the initial phonon number $N$ is conserved. Instead, if the decay of the membrane is considered the total phonon number becomes time-dependent and follows the approximate exponential decay $N(t) \approx N_0 \exp[-2\gamma t]$ (assuming equal  decay rates of the two mechanical modes, $\gamma_1=\gamma_2=\gamma$). Consequently, the two critical couplings $g_{c,i}(t)$ (with $i=1,2$) increase with time. The value of $g_{c,2}(t)$ is most important here for the stability of DTC order and Eq.~(\ref{criticalp}) gives:
\begin{equation}
g_{c,2}(t) = \sqrt{\frac{(\Delta_2^{2}+\kappa^{2})J}{4\Delta_2|\alpha|^{2}N_0}} \exp[\gamma t].
\end{equation}
As seen in Fig.~\ref{phasediag1}, DTC order only occurs for $g > g_{c,2}$. Otherwise, the relaxation phase drives the system to the normal state and the DTC order cannot persist. Imposing $g_{c,2}(t)=g$, the lifetime $T_0$ of the DTC is found as follows   
\begin{equation}
T_0=\frac{1}{\gamma}\ln\left[\frac{g}{g_{c,2}(0)}\right]. \label{t0}
\end{equation}
For $t<T_0$, the amplitudes of the period-doubling oscillations decay as:
\begin{align}
\bar{d}(t)&=\frac{2g|\alpha|N_0\exp[-2\gamma t]}{\Delta_2-i\kappa}\sqrt{1-\frac{g_{c,2}^4(t)}{g^4}}, \nonumber\\
\delta \bar{N}(t) & =  \frac{N_0\exp[-2\gamma t]}{2}\sqrt{1-\frac{g_{c,2}^{4}(t)}{g^{4}}},\label{dnt}
\end{align}
while for $t\geq T_0$ the period-doubling oscillations have disappeared. An example of DTC dynamics with finite mechanical damping is shown in Fig.~\ref{DTC-gammadecay}, finding excellent agreement with Eqs.~(\ref{t0}) and~(\ref{dnt}).

\section{Summary}\label{Sec: Summary}
In this work we have proposed a scheme for the realization of a superradiant-type quantum phase transition in a ``two membranes in the middle'' optomechanical system. We identify a regime in which the system can be accurately mapped to the Dicke model and exhibits $Z_2$ symmetry breaking in the thermodynamic limit. By modulating the drive amplitude and detuning in a periodic way, making the system cross the normal/superradiant critical point, one can realize a discrete time crystal order with period doubling. We show that such period doubling is robust to parameter deviations and persists in the thermodynamic limit.

\acknowledgments
S.C. acknowledges support from the Innovation Program for Quantum Science and Technology (Grant No. 2021ZD0301602), the National Science Association Funds (Grant No. U2230402), and the National Natural Science Foundation of China (Grant Nos. 11974040 and 12150610464).

\appendix
\section{Feasibility of the model}\label{Appendixa}

\begin{figure}
\begin{centering}
\includegraphics[scale=0.4]{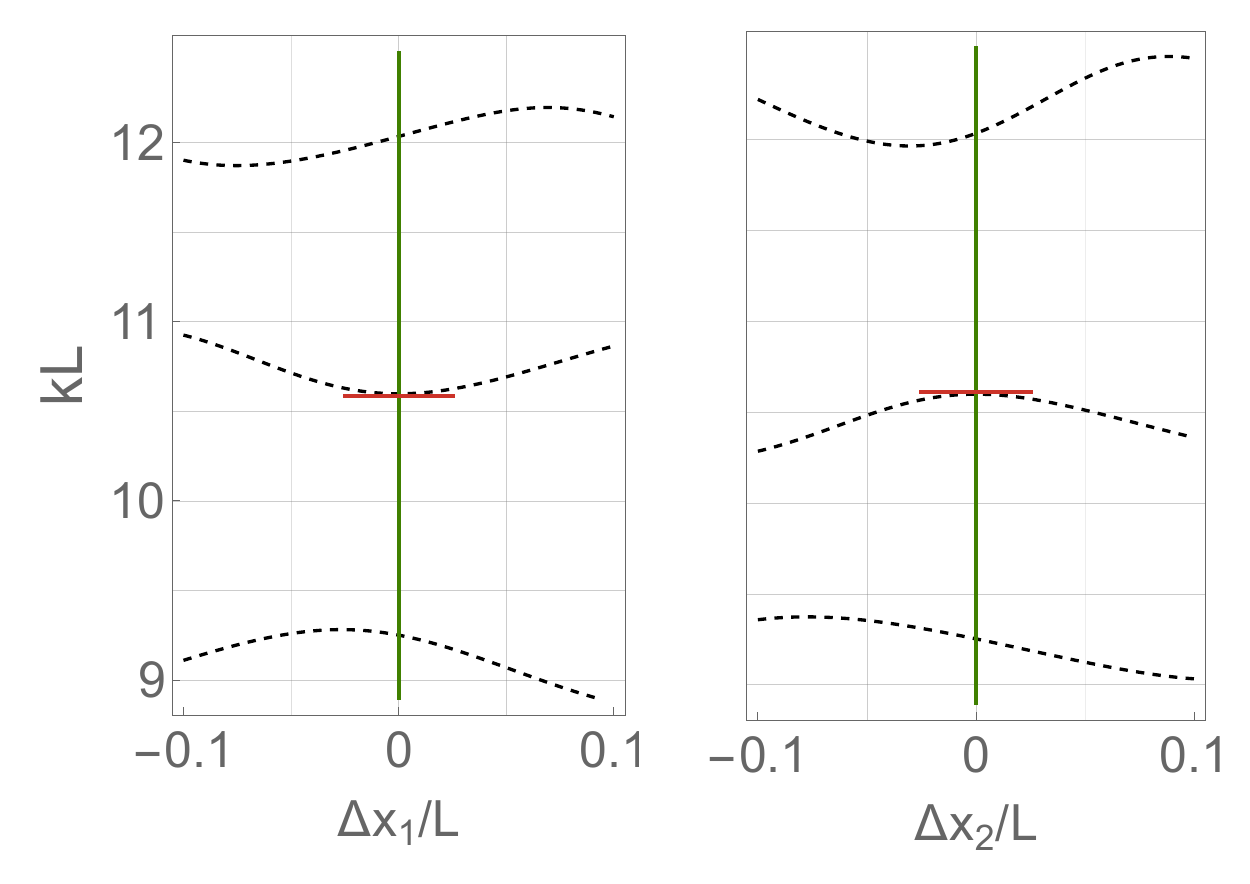}
\par\end{centering}
\caption{\label{fig:AFig-1}
Optical spectrum of the two-membrane cavity, close to the condition determined by Eq.~(\ref{eq:equilibriumpos}). Here we choose $m_{0}=7,m_{1}=-1,m_{2}=1$ and $\mathcal{T}=0.85$, giving $x^{(0)}_{1}/L\simeq-0.296$ and $x_{2}^{(0)}/L\simeq 0.407$. The left (right) panel shows the optical spectrum as function of $\Delta x_1$ ($\Delta x_2$), while setting $\Delta {x_2}=0$ ($\Delta {x_1}=0$). For the middle mode, the vanishing derivatives (indicated by red lines) imply zero first-order optomechancial couplings. The opposite curvatures with respect to changes in $\Delta x_{1}$ and $\Delta x_{2}$ correspond to second-order optomechancial couplings with opposite sign.}
\end{figure}

\begin{figure}
\begin{centering}
\includegraphics[scale=0.6]{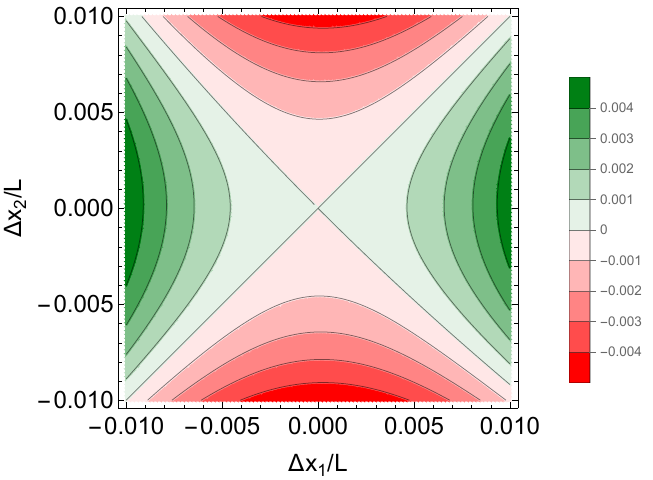}
\par\end{centering}
\caption{\label{fig:Afig-2} Optical spectrum for the middle mode of Fig.~\ref{fig:AFig-1}. More precisely, we plot $L\Delta k=L[ k(x_1,x_2)-k(x^{(0)}_1,x^{(0)}_2)]$ as function of $\Delta x_1, \Delta x_2$which. As expected, $\Delta k$ is approximately $\propto \Delta x_1^2 - \Delta x_2^2$ for small values of  $\Delta x_{1,2}$.}
\end{figure}

In this appendix, we demonstrate an implementation of the model. The
specific example we consider is a Fabry-Perot cavity with two
membranes in the middle. As indicated in Fig. 1, the laser axis is
along the $x$ direction, with the two end mirrors at positions $x=\pm L$.
In analogy to Refs.~\cite{Spencer19722_PRA5-893,Bhattacharya2008_PRA78-041801,Hartmann2008_PRL101-200503},
we model the membranes as dielectric \textquotedblleft bumps\textquotedblright
with transmission coefficient $\mathcal{T}$ and apply suitable boundary
conditions at the end mirrors and at the positions of the two membranes. A transcendental equation
can be obtained
\begin{align}\label{eq:spectrum}
\sin(2kL+2\varphi)+\sin(2kL+2k(x_{1}-x_{2}))\sin^{2}\varphi\\
-2\sin\varphi\cos(k(x_{1}-x_{2})-\varphi)\cos(k(x_{1}+x_{2})) & =0,\nonumber 
\end{align}
where $k$ is the wave number of the optical mode and $\varphi=\arccos\sqrt{\mathcal{T}}$. We now impose the condition of vanishing first-order optomechanical couplings, and require that the second-order couplings have the same form of Eq.~(\ref{HI}):
\begin{align}\label{eq:conditions_dkdx}
\frac{\partial k}{\partial x_1}=\frac{\partial k}{\partial x_2}= \frac{\partial^2 k}{\partial x_1 \partial x_2}=0, \qquad
\frac{\partial^2 k}{\partial x_1^2 }= -\frac{\partial^2 k}{\partial x_2^2 }.
\end{align}
By evaluating the first and second-order derivatives of Eq.~(\ref{eq:spectrum}), we rewrite Eq.~(\ref{eq:conditions_dkdx}) as:
\begin{align}\label{eq:feasilitycond}
&\sin(2kx_{1}-\varphi)+\cos(2k(L+x_{1}-x_{2}))\sin\varphi  =0,\nonumber \\
&\sin(2kx_{2}+\varphi)-\cos(2k(L+x_{1}-x_{2}))\sin\varphi  =0,\nonumber \\
&\sin(2k(L+x_{1}-x_{2}))  =0 \nonumber \\
&\cos(2kx_{1}-\varphi) = -\cos(2kx_{2}+\varphi) .
\end{align}
A solution of Eq.~(\ref{eq:feasilitycond}) is:
\begin{align}
k & =\frac{(2m_{0}+1)\pi}{2L}-\frac{\varphi}{L},\nonumber \\
x_{1}^{(0)} & =\frac{m_{1}\pi}{k},\nonumber \\
x_{2}^{(0)}  & =\frac{m_{2}\pi+\pi/2-\varphi}{k}, \label{eq:equilibriumpos}
\end{align}
where $m_0,m_1,m_2$ are integers. For specific parameters, the dependence of the optical spectrum as function of the displacements $\Delta x_i = x_{i}-x_{i}^{(0)}$ ($i=1,2$) is illustrated in Figs.~\ref{fig:AFig-1} and \ref{fig:Afig-2}, showing that one of the eigenmodes follows the expected behavior.

\section{Phase transition from the mean-field equations}\label{Appendixb}
The effective Hamiltonian is shown in Eq.~(\ref{effectiveH}) and the corresponding mean-field equations of motion are presented in Eq.~(\ref{langevinequation}). By substituting $b_{1,2}\rightarrow \beta_{1,2}e^{i\omega t}$, where $\beta_{1,2}$ are time-independent amplitudes, we get:
\begin{align}
\omega \beta_{1}& =  -2g|\alpha| (d+d^{*})\beta_{1}+2J\beta_{2}\nonumber,\\
\omega \beta_{2} & =  2g|\alpha| (d+d^{*})\beta_{2}+2J\beta_{1}.
\end{align}
which can only have a nontrivial solution for
\begin{equation}\label{omega_effective}
\omega = \pm\sqrt{4g^{2}|\alpha|^{2}(d+d^{*})^{2}+4J^{2}}.
\end{equation}
This equation describes how the unperturbed frequencies $\pm 2J$ of the normal modes are modified by a finite expectation value of the cavity field. In terms of $\omega$ and $d$, and by using the conserved total phonon number $N=|b_1|^2+|b_2|^2$, we find: 
\begin{align}
&|\beta_1|^{2}  =\frac{4J^{2}N}{4J^{2}+[2g|\alpha|(d+d^{*})+\omega]^{2}}, \nonumber\\
& |\beta_{2}|^2  = \frac{N[2g|\alpha| (d+d^{*})+\omega]^2}{4J^{2}+[2g|\alpha|(d+d^{*})+\omega]^{2}}. \label{beta12}
\end{align}
Assuming a stationary $d$, the mean-field equations also give $d = 2g|\alpha|(|\beta_2|^2-|\beta_1|^2)/(\Delta-i\kappa)$, implying:
\begin{align}
(\Delta^2+\kappa^2)(d+d^*)= \pm\frac{4g^2|\alpha|^2\Delta N(d+d^{*})}{\sqrt{J^{2}+g^{2}(d+d^{*})^{2}|\alpha|^{2}}},
\end{align}
where the $\pm$ sign corresponds to Eq.~(\ref{omega_effective}) and, assuming $d \neq 0$, must be chosen in accordance with the sign of $\Delta$. For $g>g_c$, the above condition has the following nontrivial solution:
\begin{align}\label{dplusddag}
(d+d^*)^{2} = \frac{16g^{2}|\alpha|^{2}\Delta N^{2}}{(\Delta^{2}+\kappa^{2})^{2}}\left(1-\frac{g_c^4}{g^4}\right),
\end{align}
which is in agreement with Eqs.~(\ref{criticalp}) and~(\ref{orderp_d}). Substituting Eq.~(\ref{dplusddag}) in the expression (\ref{omega_effective}) for $\omega$, we find the effective frequency (\ref{effective_omega}) given in the main text. Finally, we can recover Eq.~(\ref{orderp_dN}) using Eq.~(\ref{beta12}). 

\section{Effective potential of the cavity}\label{Appendixd}
We first rewrite the mean-field equation for the cavity, see Eq.~(\ref{langevinequation}), in terms of the quadrature variables:
\begin{align}
&\frac{dx}{dt}=\Delta p -\kappa x, \nonumber \\
&\frac{dp}{dt}=-\Delta x -2\sqrt{2}g|\alpha|(|b_1|^2-|b_2|^2)-\kappa p, 
\end{align}
where  $x=(d^*+d)/\sqrt{2}$ and $p=i(d^*-d)/\sqrt{2}$. It is then simple to derive the following equation of motion for $x$:
\begin{align}
\frac{d^2x}{dt^2}=-\frac{dV_{\rm eff}(x)}{dx}-2\kappa \frac{dx}{dt}.
\end{align}
where $\frac{dV_{\rm eff}(x)}{dx}= (\Delta^2+\kappa^2)x+2\sqrt{2}g |\alpha|\Delta (|b_1|^2-|b_2|^2)$. With the values of $|b_{1,2}|^2$ given by Eqs.~(\ref{beta12}) and (\ref{omega_effective}), the following explicit form of the effective potential is found:
\begin{equation}\label{Veff}
V_{\rm eff}(x) = \frac12(\Delta^2 +\kappa^2)x^2 + 2\Delta N\sqrt{J^2+2|\alpha|^2 g^2 x^2}.
\end{equation}
The critical coupling $g$ associated with $V_{\rm eff}(x)$ coincides with Eq.~(\ref{criticalp}), and in the broken-symmetry phase the minima of $V_{\rm eff}(x)$ agree with Eq.~(\ref{orderp_d}).

\bibliography{ref-superradiance}


\end{document}


\section*{Appendix A}

In this appendix, we demonstrate an implementation of the model. The
specific example we consider here is a Fabry-Perot cavity with two
membranes in the middle. As indicated in Fig. 1, the laser axis is
along the $x$ direction with the end mirrors at $x=L$ and $x=-L$.
In analogy to \cite{Spencer19722_PRA5-893,Bhattacharya2008_PRA78-041801,Hartmann2008_PRL101-200503},
we also model the membranes as dielectric \textquotedblleft bumps\textquotedblright{}
with transmission coefficient $\mathcal{T}$ and consider the boundary
conditions at the mirrors and membranes. A transcendental equation
can be obtained,
\begin{align}
\sin(2kL+2\varphi)+\sin(2kL+2k(x_{1}-x_{2}))\sin^{2}\varphi\label{eq:spectrum}\\
-2\sin\varphi\cos(k(x_{1}-x_{2})-\varphi)\cos(k(x_{1}+x_{2})) & =0.\nonumber 
\end{align}
Here, $k$ is the wave number of the optical mode, $\varphi=\arccos\sqrt{\mathcal{T}}$.

We focus on finding the appropriate position $x_{1}$ and $x_{2}$
for the two membranes that satisfy the disappeared first-order optomechanical
couplings and the opposite second-order optomechanical couplings for
the two membranes, which requires
\begin{align}
k^{(1,0)}[x_{1},x_{2}] & =0\nonumber \\
k^{(0,1)}[x_{1},x_{2}] & =0\nonumber \\
k^{(2,0)}[x_{1},x_{2}] & =-k^{(0,2)}[x_{1},x_{2}]\nonumber \\
k^{(1,1)}[x_{1},x_{2}] & =0\label{eq:feasibility}
\end{align}
We take the first and second-order derivatives of both sides of Eq.
(\ref{eq:spectrum}) and get the expression of $k^{(1,0)}[x_{1},x_{2}],k^{(0,1)}[x_{1},x_{2}],k^{(2,0)}[x_{1},x_{2}],k^{(0,2)}[x_{1},x_{2}]$
and $k^{(1,1)}[x_{1},x_{2}]$. The conditions in Eq. (\ref{eq:feasibility})
become
\begin{align}
\sin(2kx_{1}-\varphi)+\cos(2k(L+x_{1}-x_{2}))\sin\varphi & =0,\nonumber \\
\sin(2kx_{2}+\varphi)-\cos(2k(L+x_{1}-x_{2}))\sin\varphi & =0,\nonumber \\
\cos(2kx_{1}-\varphi)-\sin(2k(L+x_{1}-x_{2}))\sin\varphi & =c,\nonumber \\
\cos(2kx_{2}+\varphi)-\sin(2k(L+x_{1}-x_{2}))\sin\varphi & =-c,\nonumber \\
\sin(2k(L+x_{1}-x_{2})) & =0.\label{eq:feasilitycond}
\end{align}
By solving the equations in Eq.(\ref{eq:feasilitycond}), we get two
similar solutions and one of them is
\begin{align}
k & =\frac{(2m_{0}+1)\pi}{2L}-\frac{\varphi}{L},\nonumber \\
x_{1} & =\frac{m_{1}\pi}{k},\nonumber \\
x_{2} & =\frac{m_{2}\pi+\pi/2-\varphi}{k}.\label{eq:equilibriumpos}
\end{align}
Here, $m_{0/1/2}$ is interger.

\begin{figure}

\begin{centering}
\includegraphics[scale=0.4]{AFig-1}
\par\end{centering}
\caption{\label{fig:AFig-1}Optical spectrum for the two-membrane cavity shown
as a function of the membrane position $\Delta x_{1}$ or $\Delta x_{2}$.
$\Delta x_{1/2}$ is the position of the first (second) membrane with
respect to its equilibrium position $x_{1/2}$ (Eq. (\ref{eq:equilibriumpos})).
The two equilibrium position are represent by the (green) vertical
lines. Here we choose $m_{0}=7,m_{1}=-1,m_{2}=1$ and $\mathcal{T}=0.85$
and the equilibrium positions for two membranes are $x_{1}/L\approx-0.296$
and $x_{2}/L=0.407$, respectively.}
\end{figure}

For a given set of parameters, the numerical solution of wave number
$k$ concerning one moving membrane is shown in Fig. (\ref{fig:AFig-1}).
The other membrane is fixed at its equilibrium position. The moving
membranes corresponding to the left and right panels are membranes
1 and 2, respectively. As shown in Fig. (\ref{fig:AFig-1}), the interaction
with disappeared first-order couplings and the opposite second-order
couplings is available.

\begin{figure}
\begin{centering}
\includegraphics[scale=0.6]{AFig-2}
\par\end{centering}
\caption{\label{fig:Afig-2}Optical spectrum $k-k_{E}$ for the two-membranes
cavity shown as a function of the membrane position $\Delta x_{1}$
and $\Delta x_{2}$. $\Delta x_{1/2}$ is the position of the first
(second) membrane with respect to its equilibrium position $x_{1/2}$
(Eq. (\ref{eq:equilibriumpos})) and $k_{E}$ is the corresponding
wave number when two membranes locate at each equilibrium position.
In the red region, $k<k_{F}$ and $k>k_{F}$ in the green region.
Here we choose $m_{0}=7,m_{1}=-1,m_{2}=1$ and $\mathcal{T}=0.85$
and the equilibrium positions for two membranes are $x_{1}/L\approx-0.296$
and $x_{2}/L=0.407$, respectively.}
\end{figure}

For a given set of parameters, the numerical solution of wave number
$k-k_{E}$ concerning two moving membranes is shown in Fig. (\ref{fig:AFig-1}).
$k_{E}$ is the corresponding wave number when two membranes locate
at each equilibrium position. As shown in Fig. (\ref{fig:Afig-2}),
the wave number at the equilibrium position ($\Delta x_{1}=\Delta x_{2}=0$)
is a minimum in the $\Delta x_{1}$ direction and a maximum value
in the $\Delta x_{2}$ direction. So the desired optomechanical interaction
with disappeared first-order couplings and the opposite second-order
couplings is available.

\newpage

\bibliographystyle{unsrt}
\bibliography{ref-appendixa}